\documentclass[aps,prc,preprint,amsmath,showpacs,superscriptaddress,nofootinbib]{revtex4-1}
\usepackage{graphicx}
\usepackage{epsfig}
\newcommand{\beqy}{\begin{eqnarray}}
\newcommand{\eeqy}{\end{eqnarray}}

\begin{document}

\title{Neutron conduction in the inner crust of a neutron star in the framework of the band theory of solids}
\author{N. Chamel}
\affiliation{Institut d'Astronomie et d'Astrophysique, CP-226, Universit\'e Libre de Bruxelles, 
1050 Brussels, Belgium}

\begin{abstract}
Even though the ``free'' neutrons in the inner crust of a neutron star are superfluid, they are still strongly coupled to 
nuclei due to non-dissipative entrainment effects. These effects have been systematically studied in all regions of the 
inner crust in the framework of the band theory of solids. Using concepts from solid-state physics, it is shown that the 
density of conduction neutrons, \textit{i.e.} neutrons that are effectively ``free'', can be much smaller than the density 
of unbound neutrons (by an order of magnitude in some layers) due to Bragg scattering. These results suggest that a revision 
of the interpretation of various observable neutron-star phenomena may be necessary.
\end{abstract}


\maketitle

Born from the catastrophic gravitational core collapse of massive stars with $M\gtrsim 8 M_\odot$ at the end 
point of their evolution during type II supernova explosions, neutron stars are among the most compact objects in the Universe.
The surface of a neutron star, which is composed mainly of iron, is generally obscured by a thin atmosphere. A few meters below 
the surface at densities above $\sim 10^4$ g$\cdot$cm$^{-3}$, matter is so compressed that atoms, which are arranged in a 
regular Coulomb lattice, are fully ionized and coexist with a degenerate electron gas. Deeper in the star, 
nuclei become more and more neutron-rich~\cite{pgc11}. At densities above $\sim 4.10^{11}$ g$\cdot$cm$^{-3}$, some 
neutrons drip out of ``nuclei'' thus forming a neutron liquid~\cite{onsi08}, which is expected to be superfluid at low enough 
temperatures~\cite{cha10}. The crust of a neutron star extends up to about $\sim 10^{14}$ g$\cdot$cm$^{-3}$ (i.e. half the 
density found inside heavy nuclei), at which point nuclei dissolve into a uniform mixture of electrons and 
nucleons. 

Various astrophysical neutron-star phenomena like pulsar glitches or quasiperiodic oscillations in soft-gamma repeaters are 
expected to be related to the dynamical properties of the neutron superfluid in the inner crust (see e.g. ~\cite{lrr} for a recent review). 
In modeling these events, the neutron superfluid is generally assumed to be flowing freely through the crust because of the 
absence of viscous drag. This assumption is actually unrealistic. 
Even though the neutron superfluid can flow without friction, it can still be entrained by the crustal nuclei~\cite{car06,pet10,cir11}. 
Unlike viscous drag, this entrainment effect is non-dissipative and therefore it persists in the superfluid phase. 
Entrainment effects in neutron-star crusts has been already estimated, but only in the shallow layers of the inner crust 
around the neutron drip density $\sim 4.7\times 10^{11}$ g$\cdot$cm$^{-3}$ and near the crust bottom at densities above 
$5 \times 10^{13}$ g$\cdot$cm$^{-3}$ using different crust models~\cite{cha05,cch05,cha06}. These results 
suggested that there might exist regions at intermediate densities where the neutron propagation could be completely suppressed 
owing to the presence of a band gap in the energy spectrum of unbound neutrons. Such crustal layers would resist the 
flow of a neutron current in the same way as an ordinary insulator resists the flow of an electric current. 

In this paper, entrainment effects are systematically studied in all regions of the inner crust of a cold non-accreting neutron star 
using a unified treatment based on the nuclear energy density functional theory (EDF), see e.g. ~\cite{bhr03} for a 
review. In particuliar, the static long-wavelength neutron current-current correlation function is computed for various densities, 
ranging from neutron drip to the crust-core transition. This function, which can be expressed in terms of an effective neutron mass 
or less ambiguously in terms of a density of conduction neutrons, is a necessary microscopic ingredient for a realistic hydrodynamical 
description of the neutron superfluid in the crust~\cite{car06,pet10}. Both the equilibrium composition of the crust and the 
current-current correlation function have been calculated consistently using the same nuclear energy density functional. 

\section{Nuclear band theory of neutron-star crusts}

Since the seminal work of Negele\&Vautherin~\cite{nv73}, the inner crust of a neutron star has been generally studied in the framework 
of the Wigner-Seitz approximation~\cite{ws33} according to which the crust is divided into a set of independent spheres centered around 
each lattice site. Each cell can thus be seen as an isolated giant ``nucleus'' for which the usual methods from nuclear physics can be 
applied. Even though this approach has been fruitful for calculating ground-state properties (at least at not too high densities, as discussed 
in Refs.~\cite{bal06,cha07}), it is inappropriate for studying the low-energy dynamics of ``free'' neutrons which are delocalized over the whole crust
like ``free'' electrons in ordinary metals. Indeed, the interactions of unbound neutrons with the crystal lattice are highly non-local thus 
leading to long-range correlations which cannot be properly taken into account in the Wigner-Seitz approximation. 

In this work, the neutron conduction in the neutron-star crust will be studied using the band theory of solids~\cite{ash76}. Although this theory has 
been very successfully employed in various systems in optics, acoustics and condensed matter physics, its application to neutron-star crusts 
is rather recent~\cite{cha05,cch05,cha06}. The band theory relies on the assumption that the solid can be treated as a perfect crystal. Although the 
crust of a real neutron star might not be a perfect crystal (see e.g. Section 3.4 of Ref.~\cite{lrr} and references therein), this is still a 
reasonable approximation for cold non-accreting neutron stars. In the rest frame of the crust, both bound and unbound neutrons are supposed to 
be described by static periodic mean-fields. In the model we employ here, these fields are generated self-consistently using the nuclear 
energy density functional theory with a semi-local functional of the Skyrme type~\cite{bhr03}. This functional is of the form
\beqy
\label{1}
E=E_{\rm kin}+E_{\rm Coul}+E_{\rm Sky}\quad ,
\eeqy
where $E_{\rm kin}$ is the kinetic energy of the sample volume, $E_{\rm Coul}$ is the Coulomb energy (using the Kohn-Sham approximation 
for the exchange part~\cite{ks65}) and $E_{\rm Sky}=\int{\rm d}^3\pmb{r}\,\mathcal{E}_{\rm Sky}(\pmb{r})$ is the nuclear energy. 
Ignoring pairing, which represents a small correction to the total energy, $E_{\rm Sky}$ is a functional of the following local densities 
and currents with $q=n,p$ for neutron, proton respectively:
\noindent (i) the number density
\beqy
\label{2}
n_q(\pmb{r}) = \sum_{\sigma=\pm 1}n_q(\pmb{r}, \sigma; \pmb{r}, \sigma)\quad ,
\eeqy
(ii) the kinetic density
\beqy
\label{3}
\tau_q(\pmb{r}) = \sum_{\sigma=\pm 1}\int\,{\rm d}^3\pmb{r^\prime}\,\delta(\pmb{r}-\pmb{r^\prime}) \pmb{\nabla}\cdot\pmb{\nabla^\prime}
n_q(\pmb{r}, \sigma; \pmb{r^\prime}, \sigma)\quad ,
\eeqy
and (iii) the spin current vector density
\beqy
\label{4}
\pmb{J_q} (\pmb{r})=-\frac{\rm i}{2}\sum_{\sigma,\sigma^\prime=\pm 1}\int\,{\rm d}^3\pmb{r^\prime}\,\delta(\pmb{r}-\pmb{r^\prime}) (\pmb{\nabla} -\pmb{\nabla^\prime})\times n_q(\pmb{r}, \sigma; \pmb{r^\prime}, \sigma^\prime)\langle\sigma^\prime|\pmb{\hat\sigma}|\sigma\rangle \quad ,
\eeqy
where $n_q(\pmb{r}, \sigma; \pmb{r^\prime}, \sigma^\prime)$ is the density matrix in coordinate space 
(denoting the spin states by $\sigma,\sigma^\prime=1,-1$ for spin up, spin down respectively). 
Introducing the isospin index $t=0,1$ for isoscalar and isovector quantities respectively\footnote{Isoscalar quantities, 
also written without any subscript, are sums over neutrons and protons (e.g. $n_0=n=n_n+n_p$) while 
isovector quantities are differences between neutrons and protons (e.g. $n_1=n_n-n_p$).}, the Skyrme 
functional $E_{\rm Sky}$ can be expressed as
\beqy
\label{5}
\mathcal{E}_{\rm Sky}=\sum_{t=0,1} C_t^n n_t^2+C_t^{\Delta n}n_t\Delta n_t+C_t^\tau n_t\tau_t
+C_t^{\nabla J} n_t\nabla\cdot\pmb{J_t}+\frac{1}{2}C_t^J \pmb{J_t}^2\quad ,
\eeqy
The $C$-coefficients are determined by fitting experimental nuclear data and/or properties of infinite 
homogeneous nuclear matter as obtained from many-body calculations. The coefficients $C_t^n$ are generally 
not constant but depend on the (isoscalar) density $n$. 
Historically the Skyrme functional was obtained from the Hartree-Fock approximation using zero-range effective 
nucleon-nucleon interactions of the kind
\beqy
\label{6}
v_{i,j} & = & 
t_0(1+x_0 P_\sigma)\delta({\pmb{r}_{ij}})
+\frac{1}{2} t_1(1+x_1 P_\sigma)\frac{1}{\hbar^2}\left[p_{ij}^2\,
\delta({\pmb{r}_{ij}}) +\delta({\pmb{r}_{ij}})\, p_{ij}^2 \right]\nonumber\\
& &+t_2(1+x_2 P_\sigma)\frac{1}{\hbar^2}\pmb{p}_{ij}\cdot\delta(\pmb{r}_{ij})\,
 \pmb{p}_{ij}
+\frac{1}{6}t_3(1+x_3 P_\sigma)n(\pmb{r})^\alpha\,\delta(\pmb{r}_{ij})\nonumber\\
& &+\frac{\rm i}{\hbar^2}W_0(\pmb{\hat\sigma_i}+\pmb{\hat\sigma_j})\cdot
\pmb{p}_{ij}\times\delta(\pmb{r}_{ij})\,\pmb{p}_{ij} 
\quad ,
\eeqy
where $\pmb{r}_{ij} = \pmb{r}_i - \pmb{r}_j$, $\pmb{r} = (\pmb{r}_i + 
\pmb{r}_j)/2$, $\pmb{p}_{ij} = - {\rm i}\hbar(\pmb{\nabla}_i-\pmb{\nabla}_j)/2$
is the relative momentum, $P_\sigma$ is the two-body spin-exchange operator. Minimizing the total 
energy $E$ for fixed numbers of neutrons and protons leads to the set of self-consistent equations 
\beqy
\label{7}
h_q(\pmb{r})\varphi_{\alpha\pmb{k}}^{(q)}(\pmb{r})=\varepsilon_{\alpha\pmb{k}}^{(q)}\varphi_{\alpha\pmb{k}}^{(q)}(\pmb{r})\, ,
\eeqy
in which the single-particle (s.p.) Hamiltonians $h_q(\pmb{r})$ can be expressed as
\beqy
\label{8}
h_q(\pmb{r})&=&-\pmb{\nabla}\cdot B_q(\pmb{r})\pmb{\nabla} + U_q(\pmb{r})-{\rm i} \pmb{W_q}(\pmb{r})\cdot\pmb{\nabla}\times\pmb{\hat \sigma}\, ,
\eeqy
with the various fields defined by 
\beqy
\label{9}
B_q(\pmb{r})=\frac{\delta E}{\delta \tau_q(\pmb{r})}, \, U_q(\pmb{r})=\frac{\delta E}{\delta n_q(\pmb{r})},\, 
\pmb{W_q}(\pmb{r})=  \frac{\delta E}{\delta \pmb{J_q}(\pmb{r})}\, .
\eeqy
These fields depend on the local densities and currents, Eqs.~(\ref{2}), (\ref{3}) and (\ref{4}), which in turn can be expressed in 
terms of occupied s.p. wavefunctions $\varphi_{\alpha\pmb{k}}^{(q)}(\pmb{r})$ (see e.g. Ref.~\cite{bhr03}). 
The periodicity of the crystal lattice means that 
\beqy
\label{10}
B_q(\pmb{r}+\pmb{\ell})=B_q(\pmb{r}) \, , 
\eeqy
\beqy
\label{11}
U_q(\pmb{r}+\pmb{\ell})=U_q(\pmb{r}) \, , 
\eeqy
\beqy
\label{12}
\pmb{W_q}(\pmb{r}+\pmb{\ell})=\pmb{W_q}(\pmb{r}) \, , 
\eeqy
for any lattice translation vector $\pmb{\ell}$. The boundary conditions to be used in Eqs.~(\ref{7})-(\ref{8}) are imposed by 
the Floquet-Bloch theorem~\cite{ash76}
\beqy
\label{13}
\varphi^{(q)}_{\alpha\pmb{k}}(\pmb{r}+\pmb{\ell})=\exp({\rm i}\, \pmb{k}\cdot\pmb{\ell})\varphi^{(q)}_{\alpha\pmb{k}}(\pmb{r})  \, , 
\eeqy
where $\pmb{k}$ is the Bloch wave vector and $\alpha$ is the band index. 

Because protons are tightly bound to nuclei, proton band structure effects are very small (i.e. the proton s.p. energies are 
essentially independent of $\pmb{k}$) and will not be discussed here. In the following, the neutron energy bands and neutron wave 
functions will thus be denoted simply as $\varepsilon_{\alpha\pmb{k}}$ and $\varphi_{\alpha\pmb{k}}(\pmb{r})$ respectively. 

\section{Conduction neutrons} 
\label{conduction}

The static long-wavelength neutron current-current correlation function determines the mass current of the neutron liquid 
induced by a change of crystal momentum, the lattice being fixed. Neutron pairing, which gives rise to neutron superfluidity, 
is expected to have a minor impact on the current-current correlation function~\cite{cchbcs05} and will thus be 
neglected for simplicity. In the ground state, all neutron s.p. states lying below the Fermi level are occupied so that the neutron liquid is 
at rest in the crust frame. Considering a small shift $\delta\pmb{k}$ in $\pmb{k}$-space of the Fermi surface (FS), each s.p. 
state will then carry on average a net crystal momentum $\pmb{p_n}\equiv\hbar\delta\pmb{k}$ leading to a net mass current 
$\pmb{j_n}$. Neglecting back flow effects (i.e. assuming that the s.p. energies remain unaffected), the current is given to 
first order in $\delta\pmb{k}$ by 
\beqy
\label{14}
\pmb{j_n}= m_n \sum_\alpha\int \frac{{\rm d}^3\pmb{k}}{(2\pi)^3\hbar}\delta\tilde{n}_{\alpha\pmb{k}}\pmb{\nabla}_{\pmb{k}} \varepsilon_{\alpha\pmb{k}}\equiv n_n^c\, \pmb{p_n}
\eeqy
where $m_n$ is the neutron mass, $\delta\tilde{n}_{\alpha\pmb{k}}\equiv \tilde{n}_{\alpha\pmb{k}}-\tilde{n}^{0}_{\alpha\pmb{k}}$ denotes the change in the 
distribution function ($\tilde{n}^{0}_{\alpha\pmb{k}}$ being the ground-state distribution)  while the density $n_n^{\rm c}$ is defined in 
terms of the trace of the effective mass tensor introduced in solid-state physics~\cite{ash76}
\beqy
\label{15}
\left(\frac{1}{m_n^*(\pmb{k})^\alpha}\right)_{ij} = \frac{1}{\hbar^2}\frac{\partial^2 \varepsilon_{\alpha\pmb{k}}}{\partial k_i\partial k_j}\, ,
\eeqy
\beqy
\label{16}
n_n^{\rm c}=\frac{1}{3}\sum_\alpha \int \frac{{\rm d}^3k}{(2\pi^3)} \tilde{n}^{0}_{\alpha\pmb{k}} {\rm Tr}\biggl[\frac{m_n}{m_n^*(\pmb{k})^\alpha}\biggr]\, .
\eeqy
Incidentally the neutron effective mass tensor~(\ref{15}) has been also introduced for the study of neutron diffraction in ordinary 
crystals~\cite{ze86,ra95}. 
Since the ground-state distribution is simply given by $\tilde{n}^{0}_{\alpha\pmb{k}}=H(\varepsilon_{\rm F}-\varepsilon_{\alpha\pmb{k}})$, 
where $H(x)$ is the Heaviside unit-step distribution, the integral in Eq.(\ref{16}) has to be taken over the Fermi volume. The Fermi energy
$\varepsilon_{\rm F}$ is the Lagrange multiplier introduced during the minimization of the total energy (\ref{1}) in order to ensure the 
conservation of the neutron number
\beqy
\label{17}
n_n=\sum_\alpha \int \frac{{\rm d}^3k}{(2\pi^3)} \tilde{n}^{0}_{\alpha\pmb{k}}\, .
\eeqy

The density $n_n^c$ can be interpreted as the density of conduction neutrons by analogy with conduction electrons in ordinary metals. 
These conduction neutrons are not entrained by nuclei and can thus be considered as being effectively ``free''. Equation~(\ref{14})
implies that in an arbitrary frame where the crust moves with a velocity $\pmb{v_p}$, the neutron mass current will no longer be aligned 
with the neutron momentum, but will be given by 
\beqy
\label{18}
\pmb{j_n}=n_n^c\, \pmb{p_n} + (n_n-n_n^c) \pmb{v_p}\, .
\eeqy
The quantity $n_n-n_n^c$ can be interpreted as the density of neutrons that are effectively bound to nuclei. 

Alternatively, the neutron conduction can be expressed in terms of an effective mass $m_n^\star$ by writing 
the neutron momentum in the crust frame
\beqy
\label{19}
\pmb{p_n}\equiv m_n^\star \pmb{v_n}\, ,
\eeqy
where $\pmb{v_n}$ is the average velocity of free neutrons defined by 
\beqy
\label{20}
\pmb{j_n}\equiv n_n^{\rm f} m_n \pmb{v_n}\, .
\eeqy 
Comparing Eqs.~(\ref{19}) and (\ref{20}) with (\ref{14}) leads to the following expression for the effective mass
\beqy
\label{21}
m_n^\star = m_n \frac{n_n^{\rm f}}{n_n^{\rm c}}\, .
\eeqy

%--------------------------------------------------------------------------------------------------------------------------------
\begin{figure}
\includegraphics[scale=0.6]{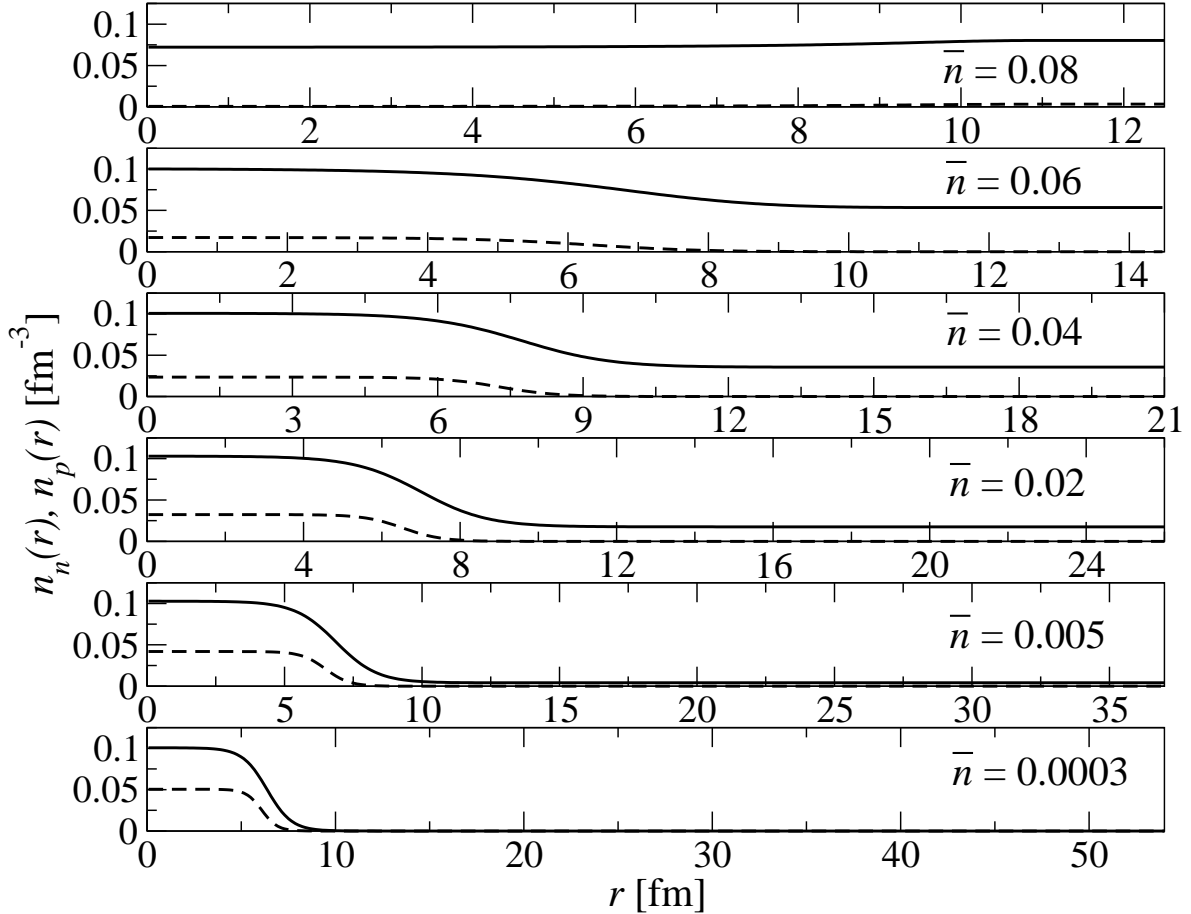}
\vskip -0.5cm
\caption{Neutron (solid line) and proton (dashed line) density profiles inside the Wigner-Seitz cell for different 
baryon densities $\bar n$ (in fm$^{-3}$), as obtained with the ETFSI method~\cite{onsi08}. Note the formation of 
``bubbles'' at $\bar n=0.08$ fm$^{-3}$: the nucleon densities are slightly larger at the cell edge than at the cell center.}
\label{fig1}
\end{figure}
%---------------------------------------------------------------------------------------------------------------------------------

\section{Numerical results}
\label{results}

In principle, the ground-state structure of the neutron-star crust at a given average nucleon density could be determined by 
solving self-consistently the EDF equations~(\ref{7}),(\ref{8}) and (\ref{9}) with Bloch boundary conditions~(\ref{13}) under the constraint 
of beta equilibrium. Considering that the equilibrium structure of the crust is a body centered cubic crystal, these calculations 
should be repeated for different lattice spacings until the lowest total energy~(\ref{1}) is found. Such calculations would be 
computionally extremely expensive. For this reason, we have taken the composition of the crust, as found in Ref.~\cite{onsi08}. 
These calculations were based on the fourth-order Extended Thomas-Fermi method with proton shell effects added via the Strutinsky-Integral 
theorem. Neutron shell effects were neglected since they were shown to be much smaller than proton shell effects~\cite{oya94}.
This so-called ETFSI method is actually a high-speed approximation to the self-consistent EDF equations~(\ref{7}),(\ref{8}) and (\ref{9}). 
The functional BSk14~\cite{sg07} used in Ref.~\cite{onsi08} was not only fitted to essentially all the available experimental atomic mass 
data with a root mean square deviation of 0.73 MeV, but was also constrained to reproduce the neutron-matter equation of state 
of Ref.~\cite{fp81}, obtained from many-body calculations using realistic two- and three-body nucleon-nucleon interactions. 
As a matter 
of fact, this equation of state is in good agreement with more recent calculations~\cite{apr98,ger10,heb10} in the density domain 
relevant to neutron-star crusts. For all these reasons, the Skyrme interaction BSk14 is particularly well-suited for studying 
neutron-star crusts. 
The neutron and proton density distributions are shown in Fig.~\ref{fig1} for a few crustal layers. 

The neutron energy bands $\varepsilon_{\alpha\pmb{k}}$ have been calculated by solving Eqs.~(\ref{7}),(\ref{8}) and (\ref{9}) with Bloch 
boundary conditions~(\ref{13}) using the self-consistent fields obtained in Ref.\cite{onsi08}. The spin-orbit potential $\pmb{W_n}(\pmb{r})$, 
which is proportional to $\pmb{\nabla} n_n(\pmb{r})$ and $\pmb{\nabla} n_p(\pmb{r})$, is much smaller in neutron-star crusts than that in 
isolated nuclei and have therefore been neglected. The neutron band structure has been computed by expanding the neutron s.p. wave functions 
into plane waves
\beqy
\label{22}
\varphi_{\pmb{k}}(\pmb{r})=\exp({\rm i} \pmb{k}\cdot\pmb{r})\sum_{\pmb{G}} \widetilde{\varphi_{\pmb{k}}}(\pmb{G}) \exp({\rm i} \pmb{G}\cdot\pmb{r})
\eeqy
where $\pmb{G}$ are reciprocal lattice vectors. By definition, $\exp({\rm i} \pmb{G}\cdot\pmb{\ell})=1$ so that the Bloch boundary conditions~(\ref{13}) 
are automatically satisfied. In this way, the EDF equation for the neutrons reduces to the matrix eigenvalue problem
\beqy
\label{23}
\sum_\beta\biggl[ (\pmb{k}+\pmb{G}_\alpha)\cdot (\pmb{k}+\pmb{G}_\beta) \widetilde{B_n}(\pmb{G}_\beta-\pmb{G}_\alpha) + \widetilde{U_n}(\pmb{G}_\beta-\pmb{G}_\alpha)\biggr] \widetilde{\varphi}_{\pmb{k}}(\pmb{G}_\beta) = \varepsilon_{\alpha\pmb{k}}\, \widetilde{\varphi}_{\pmb{k}}(\pmb{G}_\alpha)
\eeqy
with
\beqy
\label{24}
 \widetilde{B_n}(\pmb{G})=\frac{1}{V_{\rm cell}}\int{\rm d}^3r\, B_n(\pmb{r}) \exp(-{\rm i} \pmb{G}\cdot\pmb{r})\, , \ \widetilde{U_n}(\pmb{G})=\frac{1}{V_{\rm cell}}\int{\rm d}^3r\, U_n(\pmb{r}) \exp(-{\rm i} \pmb{G}\cdot\pmb{r})\, .
\eeqy
These integrals, taken over any primitive cell of volume $V_{\rm cell}$, can be efficiently calculated using fast Fourier transforms. 

%--------------------------------------------------------------------------------------------------------------------------------
\begin{figure}
\includegraphics[scale=0.6]{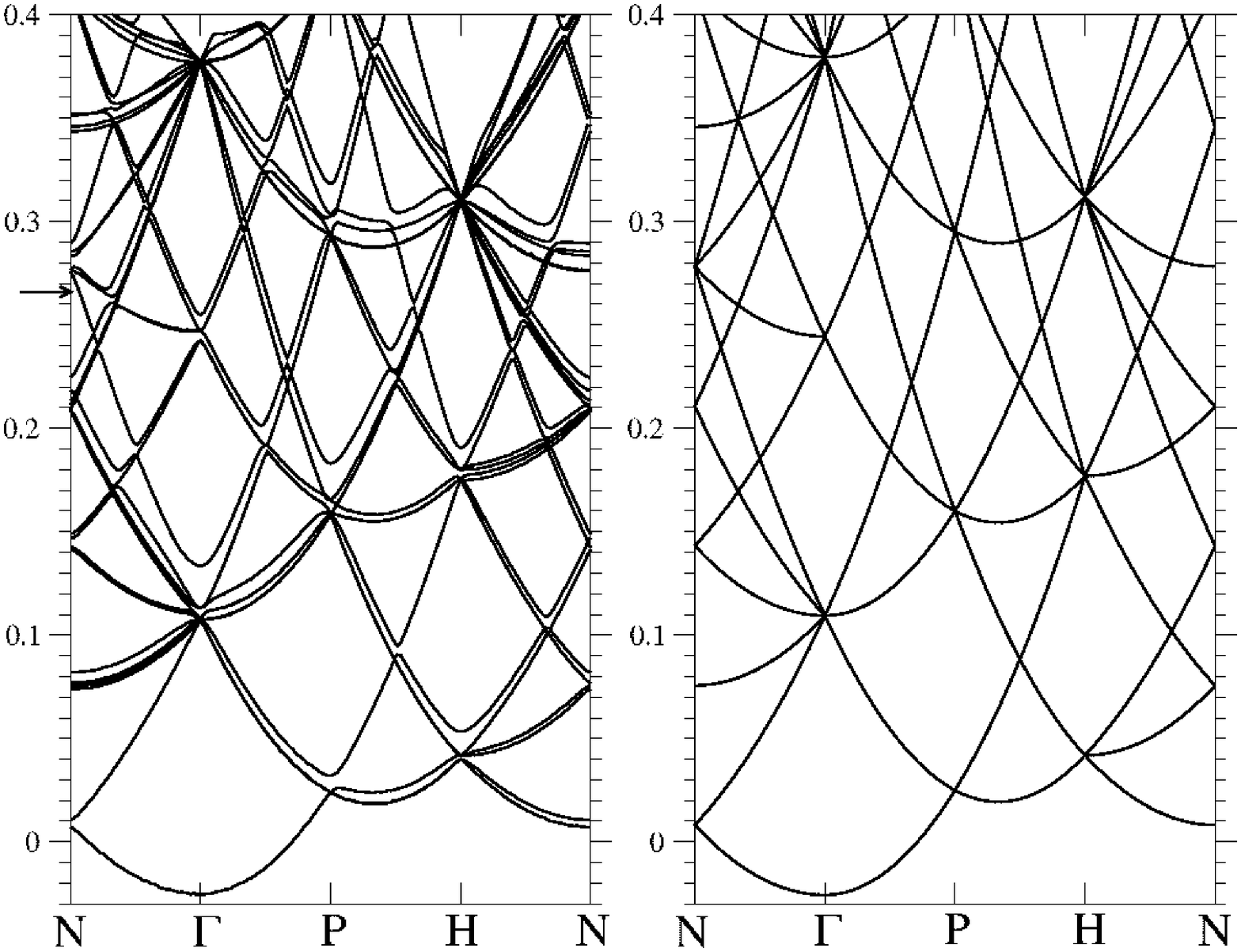}
\vskip -0.5cm
\caption{Left panel: neutron band structure in the inner crust of a neutron star at the average baryon density $\bar n=0.0003$ fm$^{-3}$
along high-symmetry lines in the first Brillouin zone (only unbound states are shown). 
The arrow indicates the position of the neutron Fermi energy. 
Right panel: band structure of a uniform neutron gas at density $n_n^{\rm f}$ (reduced zone scheme). 
For comparison with the left panel, all bands have been slightly shifted. }
\label{fig2}
\end{figure}
\begin{figure}
\includegraphics[scale=0.6]{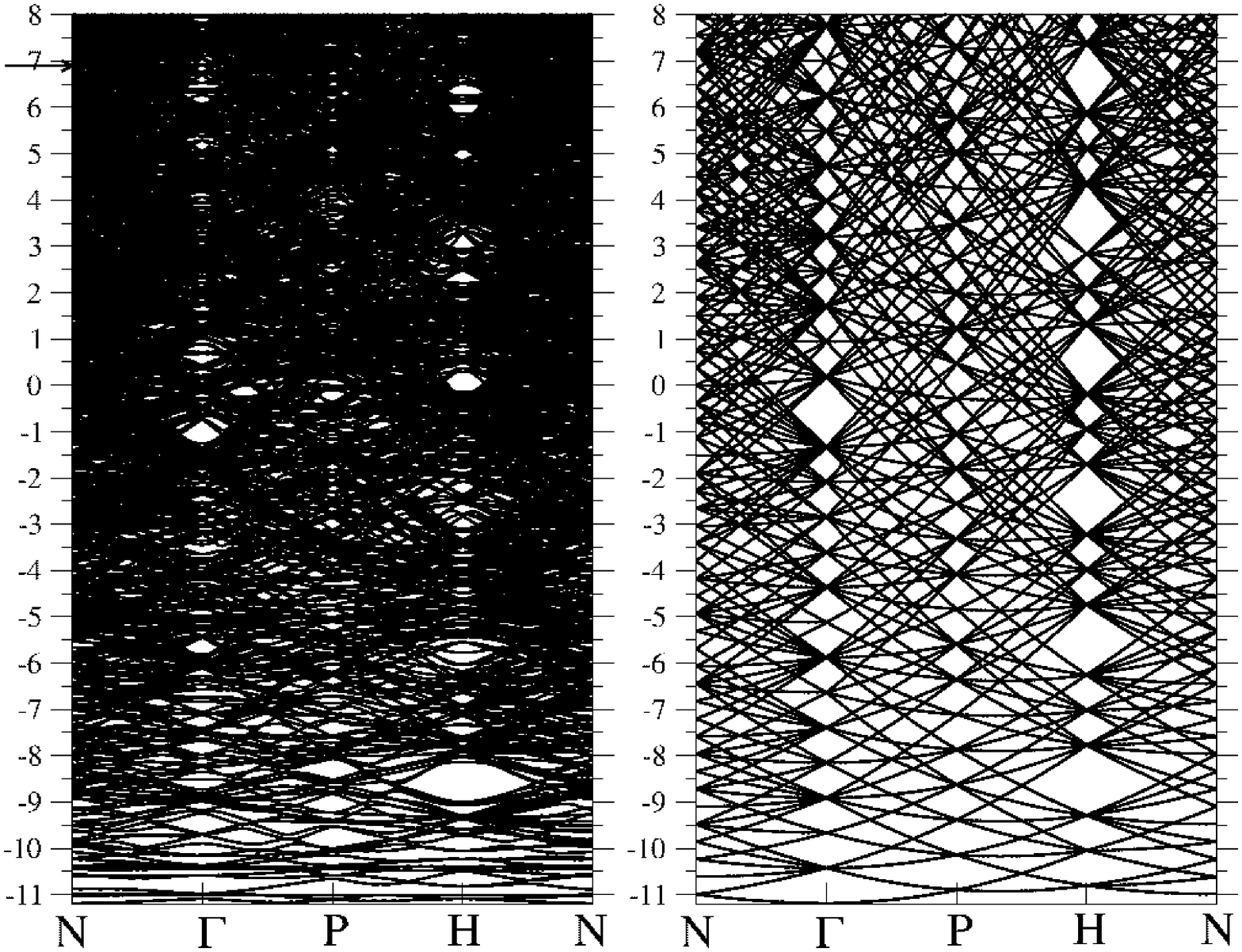}
\vskip -0.5cm
\caption{Left panel: neutron band structure in the inner crust of a neutron star at the average baryon density $\bar n=0.03$ fm$^{-3}$
along high-symmetry lines in the first Brillouin zone (only unbound states are shown). 
The arrow indicates the position of the neutron Fermi energy. 
Right panel: band structure of a uniform neutron gas at density $n_n^{\rm f}$ (reduced zone scheme).
For comparison with the left panel, all bands have been slightly shifted. }
\label{fig3}
\end{figure}
\begin{figure}
\includegraphics[scale=0.6]{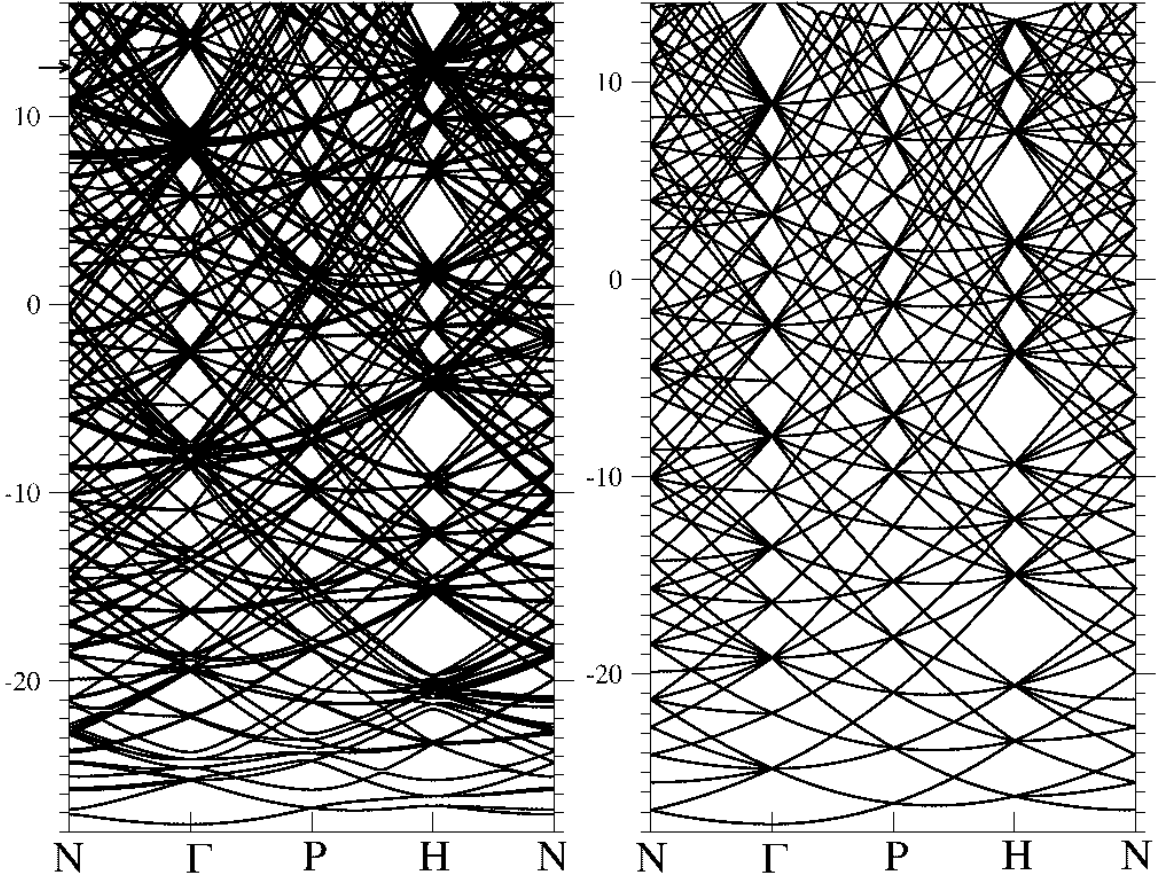}
\vskip -0.5cm
\caption{Left panel: neutron band structure in the inner crust of a neutron star at the average baryon density $\bar n=0.08$ fm$^{-3}$
along high-symmetry lines in the first Brillouin zone (only unbound states are shown). 
The arrow indicates the position of the neutron Fermi energy. 
Right panel: band structure of a uniform neutron gas at density $n_n^{\rm f}$ (reduced zone scheme).
For comparison with the left panel, all bands have been slightly shifted. }
\label{fig4}
\end{figure}
%---------------------------------------------------------------------------------------------------------------------------------

In the shallow layers of the inner crust, the linearised augmented plane wave method~\cite{cha06} would 
have been computionally much faster. Unfortunately this method can be reliably applied only in the vicinity of the neutron drip 
transition where only a few neutron bands (in the continuum) are filled. For this reason, the plane wave method has been used in all regions 
of the inner crust. 

The number of calculated neutron bands varies from a hundred in the shallowest layers of the inner crust at $\bar n=0.0003$ up to 
$850$ at $\bar n=0.03$~fm$^{-3}$. A few band structures are shown in Figs~\ref{fig2},\ref{fig3} and \ref{fig4}. We have also shown 
the band structure in the reduced zone scheme~\cite{ash76} of a uniform neutron gas of density $n_n^{\rm f}$ (empty lattice limit), 
whose energies are simply given by 
\beqy\label{25}
\varepsilon_{\alpha\pmb{k}}=B_n(n_n^{\rm f}) (\pmb{k}+\pmb{G}_\alpha)^2+U_n(n_n^{\rm f})\, .
\eeqy
As can be seen in Fig~\ref{fig2}, the energy spectrum of unbound neutron states near the neutron drip point very closely ressembles 
that of a uniform neutron gas, even though the periodic mean-field potential drops by $\sim 50$ MeV inside nuclei. This striking 
result is the consequence of the Phillips-Kleinman cancellation theorem~\cite{pk59}. The orthogonalization of the unbound states to 
the bound states leads to an effective repulsive non-local and energy-dependent potential which partially cancels the strongly 
attractive mean-field potential of nuclei. To a large extent, the same kind of cancellations occurs in the densest regions of the 
crust as can be inferred from Fig~\ref{fig4}. On the contrary, the neutron band structures in the intermediate layers of the inner 
crust differ significantly from that of a uniform neutron gas, as illustrated in Fig~\ref{fig3}. In particular, the energy spectrum 
is much denser thus revealing that on average the energies $\varepsilon_{\alpha\pmb{k}}$ have a much weaker $\pmb{k}$-dependence 
(almost flat bands) than that given by Eq.~(\ref{25}). As a result, the conduction neutron density $n_n^{\rm c}$ given by 
Eq.~(\ref{16}) is expected to be much smaller than the density $n_n^{\rm f}$ of unbound neutrons at densities $\bar n\sim 0.03$ fm$^{-3}$. 

The computation of the conduction neutron density requires the evaluation of the effective mass tensor~(\ref{15}) for all occupied
 bands. In fact, it follows from Eq.~(\ref{16}) and the periodicity of the s.p. energies that completely filled bands do 
not contribute to the current. This can be easily seen by simply rewriting Eq.~(\ref{16}) as an integral over the Fermi surface
using the Green-Ostrogradsky's theorem
\beqy
\label{26}
n_n^{\rm c}=\frac{m_n}{24\pi^3\hbar^2}\sum_\alpha \int_{\rm F} |\pmb{\nabla}_{\pmb{k}} 
\varepsilon_{\alpha\pmb{k}}|{\rm d}{\cal S}^{(\alpha)}\, .
\eeqy
Even though this expression is fully equivalent to Eq.~(\ref{16}), it is computationally much more convenient since 
only the evaluation of the first derivative of $\varepsilon_{\alpha\pmb{k}}$ is needed. In addition, these derivatives 
can be easily calculated analytically using the Hellmann-Feynman theorem~\cite{hf39}
\beqy
\label{27}
\frac{\partial\varepsilon_{\pmb{k}}}{\partial k_i}=\sum_{\pmb{G}_\alpha,\pmb{G}_\beta} \widetilde{\varphi}_{\pmb{k}}(\pmb{G}_\alpha)^*\, 
\widetilde{B_n}(\pmb{G}_\beta-\pmb{G}_\alpha)(2 k^i+G_\alpha^i+G_\beta^i)\, \widetilde{\varphi}_{\pmb{k}}(\pmb{G}_\beta)
\eeqy
with the wavefunctions normalized as
\beqy
\label{28}
\sum_\beta |\widetilde{\varphi_{\pmb{k}}}(\pmb{G}_\beta)|^2=1\, .
\eeqy
For each average density $\bar n$, the neutron Fermi energy $\varepsilon_{\rm F}$ has been determined solving Eq.~(\ref{17}) 
using the mean-value point method~\cite{bal73}. The Fermi surface integral in Eq.~(\ref{18}) has been evaluated with the 
Gilat-Raubenheimer method~\cite{gr66} using up to 1360 points in the irreducible domain (i.e. 65280 points in the first Brillouin 
zone) in order to ensure a precision of a few percent. Results are summarized in Table~\ref{tab1}. As expected from the band structures, 
the flow of neutrons is almost unaffected by nuclei in the peripheral regions of the inner crust. On the contrary, the neutron conduction 
is found to be almost completely suppressed at densities $\bar n\sim 0.02-0.03$~fm$^{-3}$. Whereas more than 90\% of neutrons are unbound 
in these layers, less than 10\% of them are actually conducting 
leading to a huge enhancement of the neutron effective mass $m_n^\star\simeq 13.6 m_n$. Incidentally, this result is in close agreement 
with the effective mass $m_n^star\simeq 15.4 m_n$ obtained in a previous work~\cite{cha05} using a different crust model thus suggesting 
that such strong entrainment effects are generic. However, further work remains to be done exploring the dependence of $m_n^\star$ on the 
nuclear energy density functional. 

\begin{table}
\begin{tabular}{|c|c|c|c|c|c|}
\hline
$\bar n$ (fm$^{-3}$) & Z & A &  $n_n^{\rm f}/n_n$ (\%) & $n_n^{\rm c}/n_n^{\rm f}$ (\%) & $m_n^\star/m_n$ \\
\hline
0.0003 &  50 & 200  & 20.0 & 82.6 & 1.21\\
0.001 & 50 & 460 & 68.6 & 27.3 & 3.66 \\
0.005 & 50 & 1140  & 86.4 & 17.5 & 5.71\\
0.01 & 40 & 1215   & 88.9 & 15.5 & 6.45\\
0.02 & 40 & 1485  & 90.3 & 7.37 & 13.6\\
0.03 & 40 & 1590  & 91.4 & 7.33 &13.6 \\
0.04 & 40 & 1610  & 88.8 & 10.6 & 9.43\\
0.05 & 20 & 800  & 91.4 & 30.0 & 3.33\\
0.06 & 20 & 780  & 91.5 & 45.9 &2.18\\
0.07 & 20 & 714 & 92.0 & 64.6 & 1.55\\
0.08 & 20 & 665 & 104 & 64.8 & 1.54 \\ 
\hline
\end{tabular}
\caption{Composition of the inner crust of cold non-accreting neutron stars as obtained from Ref.~\cite{onsi08}. 
$Z$ and $A$ are respectively the average number of protons and the \emph{total} average number of nucleons inside the 
Wigner-Seitz cell. $n_n$ is the average neutron density, $n_n^{\rm f}$ is the density of free neutrons as defined by the 
quantity $\rho_{Bn}$ in Ref.~\cite{onsi08}, $n_n^{\rm c}$ is the density of conduction neutrons and $m_n^\star$ the neutron 
effective mass. 
Note that in the densest layer, $n_n^{\rm f}>n_n$ due to the formation of bubbles as indicated in Fig.~\ref{fig1}.} 
\label{tab1}
\end{table}

\section{Microscopic origin of entrainment}

The large discrepancy between the density of unbound neutrons and the density of conducting neutrons is somehow counterintuitive. 
Indeed in ordinary metals, the electrons that are tighly bound inside the individual atoms constituting the solid have their wavefunction vanishing 
exponentially outside atoms and are therefore not much affected by the Bloch boundary conditions. As a consequence, their energy bands in 
$\pmb{k}$-space are essentially flat so that $\pmb{\nabla}_{\pmb{k}} \varepsilon_{\alpha\pmb{k}}\simeq 0$ hence yielding a negligible 
contribution to the current. The non-trivial electron band structure arises from the most loosely bound ``valence'' electrons in the isolated 
atoms which become delocalized in a metal and which can be generally identified with the conduction electrons (still, the density of valence 
electrons is not exactly equal to the density of conduction electrons). On the contrary, the 
neutron-saturated ``nuclei'' found in the inner crust of a neutron star only exist because of the Pauli blocking effect from the surrounding 
neutron liquid but would decay immediately in vacuum. For the reasons mentioned above, neutrons bound inside nuclei do not contribute 
to the current. Since unbound neutrons are delocalized, one might naively expect that they are all conducting. Indeed, ignoring the crystal 
lattice and treating the unbound neutrons as a uniform gas of density $n_n^{\rm f}$, it follows immediately from Eq.~(\ref{16}) or Eq.~(\ref{26}) 
that $n_n^{\rm c}=n_n^{\rm f}$. However it should be emphasized that the density of conduction neutrons is fundamentally different 
from the density of unbound neutrons: the former characterizes the dynamics of the neutron liquid while the latter is a ground-state property. 
These two densities are generally not equal because unbound neutrons can be scattered by the crystal according to Bragg's law. 

The effects of Bragg scattering are embeddied in the effective mass tensor~(\ref{15}) appearing in the definition~(\ref{16}) of the conduction neutron 
density. The components of this tensor need not be positive and can actually be negative for wave vectors $\pmb{k}$ such that Bragg reflection 
is allowed. In particular, an unbound neutron will be reflected whenever its Bloch wave vector $\pmb{k}$ lies on a Bragg plane, \textit{i.e.} 
$\pmb{k}$ satisfies the von Laue condition $2 \pmb{k}\cdot\pmb{G} = G^2$ for any reciprocal lattice vector $\pmb{G}$. As a result, the neutron 
current will be suppressed however small the periodic mean-field potential is. Using perturbation theory and setting $B_n(\pmb{r})\equiv B_n$ 
for simplicity, it can be shown that in the vicinity of a Bragg plane the s.p. energies of free neutrons are split into two bands~\cite{ash76}
\beqy
\label{29}
\varepsilon^{\pm}_{\pmb{k}}\simeq \frac{1}{2} B_n (k^2 + (\pmb{k}-\pmb{G})^2) \pm \sqrt{\frac{1}{4}B_n^2 (k^2 - (\pmb{k}-\pmb{G})^2)^2+\vert\widetilde{U_n}(\pmb{G})\vert^2} \, .
\eeqy
It is easily seen that the effect of the periodic potential is to flatten the bands around Bragg planes by introducing a gap
of magnitude $2\vert\widetilde{U_n}(\pmb{G})\vert$. As a consequence, the conduction neutron density will be reduced whenever 
the Fermi energy lies on the top (bottom) of the lower (upper) band. The neutron conduction thus depends on two factors: 
i) the amount of flattening of the bands which in turn is governed by the periodic potential, ii) the position 
of the Fermi level which is determined by the total neutron density. The more the Fermi sphere intersects Bragg planes, the larger 
will be the effect of Bragg scattering on the neutron conductivity. The number of intersections depends on the ratio between the Fermi 
volume and the volume of the first Brillouin zone. According to Luttinger's theorem~\cite{lut60}, the Fermi volume associated with unbound states
is given by $V_{\rm F}=(2\pi)^3 n_n^{\rm f}$. 
By definition, the volume of the first Brillouin zone is equal to $V_{\rm BZ}=(2\pi)^3/V_{\rm cell}$. Therefore the ratio 
$V_{\rm F}/V_{\rm BZ}=n_n^{\rm f}V_{\rm cell}$ is simply equal to the average number of unbound neutrons inside the Wigner-Seitz 
cell (or equivalently to the average number of unbound neutrons per nucleus). Basically, this number is the lowest at the neutron drip point, 
peaks at about $1417$ at density $\bar n=0.03$ fm$^{-3}$ and decreases at higher densities. As expected, the resistance 
of the crust to the flow of a neutron current follows a similar behavior (see Table~\ref{tab1}). 

\section{Conclusions}

Despite the absence of viscous drag, the neutron superfluid permeating the inner crust of a neutron star is still strongly 
coupled to nuclei due to non-dissipative entrainment effects. 
These effects have been systematically studied in all regions of the inner crust of a cold non-accreting neutron star in the 
framework of the band theory of solids using an effective Skyrme nuclear energy density functional which not only yields an 
excellent fit to essentially all experimental atomic masses but was also constrained to reproduce a realistic neutron-matter 
equation of state. 

Some regions of the inner crust are found to strongly resist the flow of a neutron current in the same way as an ordinary 
insulator resists the flow of an electric current. As a consequence the density of conduction neutrons, \textit{i.e.}
neutrons not entrained by nuclei, turns out to be much smaller than the density of unbound neutrons thus leading to a huge 
enhancement of the neutron effective mass. These results suggest that a revision of the interpretation of many observable 
astrophysical phenomena like pulsar glitches~\cite{cc06}, quasiperiodic oscillations in Soft-Gamma Repeaters~\cite{sam09} and 
the cooling of neutron stars~\cite{pr12}, may be necessary. 

Incidentally, the kind of entrainment effects discussed here are also expected to be observed in laboratory superfluid systems. 
Indeed, qualitatively similar results have been predicted for a unitary Fermi gas in a 1D optical lattice~\cite{wa08}. 
Since a dilute neutron gas is approximately in the unitary regime~\cite{sp05}, cold atoms experiments could shed light on the properties
of superfluid neutrons in neutron-star crusts.

\begin{acknowledgments}
The author thanks C.J. Pethick and S. Reddy for fruitful discussions. 
This work was financially supported by FNRS (Belgium)
and CompStar, a Research Networking Programme of the European Science Foundation. 
The author thanks the Institute for Nuclear Theory at the University of Washington for its hospitality 
and the Department of Energy for partial support.
\end{acknowledgments}

\end{document}